arXiv e-Print Working Paper

# Pareto-Nash Allocations under Incomplete Information: A Model of Stable Optima

Mayaki, Alfred Anate Bodurin[1]

Royal Economic Society (UK)
Open University Business School, The Open University (UK)

## Abstract

Prior literature on two-firm, two-market, and two-stage extended dynamic models has introduced what Güth (2016) succinctly terms a *social dilemma*—a state in which conglomerate firms competing in a Bertrand duopoly consider jointly optimising profits under a tacit, self-enforcing agreement to deter market entry. This theoretical article reinterprets the social dilemma highlighted by Güth (2016) not only in the context of allocation but also through the lens of competition, where entry must legally be permitted even if cooperative signalling (Mayaki, 2024) would otherwise sustain joint profitability. This study explores the significance of a sufficiency condition on each firm's non-instantaneous reaction function, requiring the maintenance of a stable long-run equilibrium through retaliative restraint, characterised by either two negative eigenvalues or a saddle-path trajectory.

## 1. Introduction

A key challenge facing regulated conglomerate downstream markets is how competition is understood and applied in practice, particularly when attempting to abridge the act of active policymaking (which is highly *reactive*) with the analytical rigour of practice-based theory (which is highly *proactive*). Well-endowed entrants still face hurdles when competing with large segment-focused conglomerates and are seldom given the option of profitability without *forbearance*, which often describes typical correspondence with incumbents by new entrants. In the present era of artificial general intelligence and complex systems and new forms of logic such as algorithmic pricing (Fainmesser, Galeotti and Momot, 2023), the corresponding challenge has become an unmistakenly prevalent empirical one of industrial organisation. There is broad consensus that forbearance (Jin and Eapen, 2022) serves as an additional pillar on how efficient conglomerate markets operate, particularly where competing multimarket firms vie for revenue gains through strategic competition. At its core, fair competition requires regulatory supervision, which should promote equitable outcomes across downstream conglomerate markets.

Therefore, when conglomerates are *in the core*—and where allocations are Pareto-optimal—it may be perceived that competition authorities and regulators have achieved their job, that is, to limit the abuse of dominant power in an already saturated environment, ideally at minimal social cost. Pricing systems can assist in defining models that accurately predict future revenue and that use practical tools such as the small but significant non-transitory change in price test used by the EU's competition authority (Mandrescu, 2018). However, as we look at theory with the accompanying stylised facts and model performance (assuming uniform pricing) and discuss implications, we find that precedent almost certainly will supersede the practical tools at hand. For example, with algorithmic pricing, the development of robust policy tools that address both horizontal and vertical mergers, discriminatory pricing, and the fostering of competitive markets, growth is as critical an endeavour as it has ever been.

[1] All correspondence and requests for information related to this article should be sent to alfred.mayaki@ou.ac.uk.

In prior studies, time-based convergence with respect to global and local optima has explained the context of growth, household utility and collective bargaining (Mayaki, 2024). However, under Güth's premise within this investigation (Güth and Saaksvuori, 2012), growth is said to carry a significance that creates what he calls the *social dilemma*. According to one paper by Güth (2016:68) what describes a social dilemma is individual rationality, which, when modelled, leads to collective irrationality. In other words, or in simpler terms, what seems best for each party acting on its own accord may result in diminished utility for all participants when all group members act similarly. Each participant has an incentive to act in their own self-interest (defect from cooperation). Güth contends that if each counterparty acts selfishly, the payoff in a Bertrand duopoly game is worse off than if they had cooperated (Güth and Saaksvuori, 2012:3). Which forms of cooperation are acceptable to Güth (2016) in his paper on social dilemmas? Just the strategic profiles that bring about what we will continue to describe as a *Pareto-Nash* core allocation. Inevitably, forbearance caveats the grim-trigger consequence in most, if not all, academic discussions pertaining to bridging research and policy in repeated games, which include aspects of game theory. In Fainmesser, Galeotti and Momot (2023), these theoretical discussions cover a series of popular models demonstrating cooperative/noncooperative, infinitely repeated/two- or three-stage elements, which can be used to facilitate what we might call stable algorithmic equilibria. Importantly, the implications of these models may be difficult to interpret without full knowledge of strategic histories, Nash equilibrium or stable optima.

This research uses empirical data from the BEIS Longitudinal Small Business Survey (LSBS), a rich cross-sectional dataset that captures business behavior across SMEs in the UK. Key variables from the LSBS inform the construction of a firm-level information asymmetry index, using proxies such as digital technology adoption, strategic awareness (e.g., competitor monitoring), advice-seeking behavior, and data capability. These proxies allow for a composite measure of asymmetric information consistent with the theoretical framework of bounded rationality and imperfect market knowledge.

With respect to the model, complementing the data sourced from the LSBS, the study draws on conceptual variables available from the University of Essex's Understanding Society (UKHLS) online repository, particularly in modelling firm growth trajectories among self-employed individuals and microenterprises. These include changes in business income, digital confidence, and business strategy, as well as subjective assessments of future expectations and uncertainty. While UKHLS is primarily household-based, its business ownership module provides valuable microlevel inputs for modelling incomplete information in small-scale firm settings. In addition, variables such as firm age, initial endowments (proxied by total assets or prior turnover), and export intensity are derived from simulated data on the basis of distributions found in Orbis (Bureau van Dijk) and the ONS Business Structure Database (BSD). These commercial and administrative datasets offer representative real-world benchmarks for financial and structural firm characteristics across UK industries. Although not directly applicable for this version of the study, they inform the calibration of inputs used in the regression specification and model validation.

By integrating these cross-sectional sources, the study captures firm-level heterogeneity in strategic behavior, particularly under varying degrees of informational advantage. The empirical design is structured to reflect the theoretical model's emphasis on mutual forbearance, export decisions, and algorithmic pricing behavior, thereby linking microeconomic firm data with macro theoretical equilibrium outcomes. This hybrid approach—combining survey-based behavioral indicators with stylised financial proxies—strengthens the study's contribution to contemporary industrial organisation and regulatory economics.

### *1.1 Methodological Note on Data Sources and Simulation*

This paper uses a combination of real-world evidence and simulated data to explore how stable, welfare-optimal outcomes can arise under conditions of incomplete information. To help shape the model's assumptions—especially around how different types of workers behave and what kinds of contracts they face—I refer to UK survey data from UKHLS and labour market statistics available through Nomis. These datasets provide insight into patterns like wage stickiness, earnings inequality, and how job security varies across regions and occupations. While they help to ground the model in real-world context, I do not use them directly for estimation or statistical testing.

Instead, to test the model's predictions and evaluate the stability of Pareto-Nash allocations, I run a series of simulations. These simulations generate artificial data based on different scenarios: for example, what happens when workers have better or worse information, when coalition costs are high or low, or when wages adjust more slowly to shocks. I then apply fixed effects regression analysis to these simulated outcomes to measure how changes in these factors affect welfare and stability. This simulation-based approach allows me to explore a wide range of conditions while staying focused on the theoretical insights of the model.

## 2. Modelling Multimarket Linkages

In the regulation of multimarket firms—often referred to as conglomerates in the literature—there is significant conceptual overlap with the notion of *forbearance* or intermarket restraint. These conglomerates, which dominate relevant markets, frequently choose to avoid retaliatory pricing or aggressive competition when operating across shared product markets. This behavior aligns with findings from studies on Bertrand duopolies, where such restraint promotes long-term strategic stability (Cracau and Sadrieh, 2019; Güth, 2016; Jin and Eapen, 2022).

Qin *et al*. (2005) examine a quantum Bertrand duopoly game with homogeneous goods under conditions of incomplete information, where firm output is governed by a downwards-inclined demand curve. Their findings suggest that *quantum entanglement*, treated as a fixed informational resource defined by parameters $N$ and $M$, can increase profits in asymmetrical settings—specifically, a larger $M$ leads to higher global profits for the firm at any given entanglement intensity $\gamma$. Similarly, Lo and Yueng (2020) explore classical and quantum versions of Stackelberg–Bertrand duopolies. They find that quantum entanglement undermines the traditional second-mover advantage, whether in scenarios with a single leader facing multiple followers or a single follower contending with multiple leaders.

## 3. Time-Based Models of Güth's Social Dilemma

Drawing upon the special case of Güth's two most recent works on social dilemmas to analyse competition approaches theoretically, the long-run equilibrium price of a competitive firm in a downstream market is optimised where eigenvalues are stable. Here, in this model, a two-firm, two-market, two-stage cooperative pricing game is grounded under the premise that conglomerate markets possess core allocation and draws upon the view that the Roche potential in a Pareto-Nash allocation within a grim-trigger *super game is* optimal only during specific phases of the game—not throughout its entire lifecycle—assuming that defection does not occur at any stage. Certain necessary conditions of optimality must be satisfied, namely, a condition referred to as a *recursive core* (Becker and Chakrabarti, 1995). Here, firms are assumed to choose between the choice of maximising short-run revenue—plus factor endowments, net of period-by-period costs—and the choice of minimising what we might refer to as a *model* efficiency wage (Akerlof and Yellen, 1985; Coad, 2007).

The extension of Güth's prior model reflects an assumption of incomplete information, where firms cannot fully observe all market conditions—firms that can observe perfect information have greater growth prospects given their drive towards profitability. This example is taken from a system of accretion disk equations used to calculate *P-type* and *S-type* orbits in a binary system (Shakura and Sunyaev, 1973), represented initially in the form of firm $i$*'s* profit function in period one of a two-stage Bertrand duopoly game. The profit for firm $i$, as noted below, is given by the objective function $f(p, q)$ where $p$ denotes the price for a differentiated good and where $q$ denotes the firm's level of output.

(3.0.1) $\pi(\dot{p}, q) = \sum_{i=1}^{n}\left(x_{i,t} + y_{i,t}\right) + \left(p_{i,t} - c_{i,t}\right) \cdot p_{i,t}\left(q_{i,t}\right) + e$

(3.0.1) is subject to the Lagrangian inequality constraint:

(3.0.2) $\pi_{i,t} \leq \lambda \cdot (\frac{1}{N}\sum_{i=1}^{N} K)$

Equation (3.0.1) consists of $x$, which signifies an initial factor endowment, and $y$, which signifies the rate of change or growth of the firm within the two-stage system. $K$ denotes the market scale or size, $\pi_t$ denotes the firm profit, and $e$ is a residual error term. Now, while we know that $x_t$ and $y_t$ are regressors, what should be said is that they are not equal. Theoretically, as per (3.0.3), $y_t$ denotes growth through asymmetrical information, whereas $x_t$ denotes a vector of time-varying observations. Available information is given as $\emptyset \in 0{,}1$, where perfect information is defined as $\emptyset = 1$ and values lower than 1 equal greater business uncertainty. Assuming that firm growth in this two-firm market occurs at a rate equal to:

(3.0.3) $y = A \cdot \sigma^{(1-\varphi)}$

The new function is defined as follows:

(3.0.3a) $\sum_{i=1}^{n}\left(x_{i,t} + A \cdot \sigma^{(1-\varphi)}\right) + \left(p_{i,t} - c_{i,t}\right) \cdot p_{i,t}\left(q_{i,t}\right) + e$

where $A$ is equal to the age of the firm (Lee, 2014), $\varphi$ denotes the extent of diminishing returns from the asymmetry of knowledge and information, and Coad's (2007) *Penrose effect* relating to price behaviour and export market growth $\sigma$ is deemed to play a significant role in this system (Coad, 2007; Fuertes-Callén and Cuellar-Fernández, 2019). Thus, pursuing *more* domestic profit ultimately harms firm $i$*'s* growth via the objective function in (3.0.3a). Güth's social dilemma only arises because of the need to balance Pareto payoffs with single-firm performance—*without* harming total surplus or compromising market stability through aggressive competition, which might attract external entrants, necessitating a strategy of forbearance. Hence, by calculating partial differentials, eigenvalues and Jacobians for Lagrangian equations (3.0.3a) *s.t.* (3.0.2) as above, this research study aims to re-express Güth's (2016) one-shot social dilemma within a two-period conglomerate duopoly price war scenario.

We first must set up the Lagrangian using the Kuhn-Tucker (KKT) multiplier used for solving the Lagrangian where an explicit inequality constraint binds the objective function. Let $\mu \geq 0$ be the KKT multiplier for the inequality constraint. The Lagrangian is:

(3.0.3b) $\mathcal{L} = \sum_{i=1}^{n}\left(x_{i,t} + A \cdot \sigma^{(1-\varphi)}\right) + \left(p_{i,t} - c_{i,t}\right) \cdot p_{i,t}\left(q_{i,t}\right) + e + \sum_{i=1}^{N} \mu_{i,t} \left[\lambda \cdot \left(\frac{1}{N}\sum_{i=1}^{N} K\right) - \pi_{i,t}\right]$

From (3.0.3b), we can derive the following first-order condition of $\mathcal{L}$ w.r.t $p$, where we assume the profit function is denoted as:

(3.0.3c) $\pi_{i,t} = \left(p_{i,t} - c_{i,t}\right)q_{i,t}$

Simplifying the quasi-concave inequality constraint $\left(\frac{1}{N}\sum_{i=1}^{N} K\right)$, which essentially confirms that domestic market profits converge to zero as $N$ increases, as represented by $\widehat{K}$, which is a condition that binds all firms.

(3.0.4) $\mathcal{L} = \sum_{i=1}^{n}\left(x_{i,t} + y_{i,t}\right) + \left(p_{i,t} - c_{i,t}\right) \cdot p_{i,t}\left(q_{i,t}\right) + \mu_{i,t}\left[\lambda\widehat{K} - \left(p_{i,t} - c_{i,t}\right)q_{i,t}\right] + e$

Taking partial derivatives for each firm *i* and setting to zero. FOCs w.r.t *p* and *q* are:

(3.0.4a) $\frac{\partial \mathcal{L}}{\partial p_{i,t}} = \left(2p_{i,t} - c_{i,t}\right)q_{i,t} + \mu_{i,t}q_{i,t} + e = 0$

(3.0.4b) $\left(2p_{i,t} - c_{i,t}\right)q_{i,t} = \mu_{i,t}q_{i,t}$

(3.0.4c) $\frac{\partial \mathcal{L}}{\partial q_{i,t}} = \left(p_{i,t} - c_{i,t}\right)p_{i,t} - \mu_{i,t}\left(p_{i,t} - c_{i,t}\right) = 0$

(3.0.4d) $\left(p_{i,t} - c_{i,t}\right)p_{i,t} = \mu_{i,t}\left(p_{i,t} - c_{i,t}\right)$

(3.0.4e) $p_{i,t} = \mu_{i,t}$

Solving for the complementary slackness condition:

(3.0.5) $\mu_{i,t} \geq 0,\ \lambda K - \pi_{i,t} \geq 1,\ \mu_{i,t} \cdot \lambda K - \pi_{i,t} = 0$

Consequently, the system found in equations (3.0.4b) and (3.0.4e) form the basis for understanding firm pricing under growth constraints and incomplete information.

### *3.1 Equilibrium Payoffs with Multimarket Competition*

Here, we explore the assumption of multimarket goods as differentiated in line with Güth (2016). The paper assumes that firm *i* wishes to optimise market *share* and not firm profit. Is there a significant difference? Sort of. Essentially, firm *i* forgoes its entitlement to excess profit at time *t* to tacitly collude with the duopolistic competitor. This tacit agreement forms the basis of what Güth (2016) refers to as *asymmetric attractiveness*. In this section, we also denote an inequality constraint $\boldsymbol{\lambda}$ forbidding algorithmic pricing beyond what we interpret as market-clearing demand. Now, if firm *j* plays an identical reaction function as that found in (3.0.1) during the second stage of this game (Mayaki, 2024), taking this example as elements within a dynamically stable system, we can simply upend the system found in Mayaki (2024:9) with backwards induction; if we work back, firm *i* retains Stackelberg's first-mover advantage in the first period.

The reason this first-mover advantage occurs is *not* only because of factor endowments or a steeper demand curve because endowments are identical, and demand is exogenous in this model. The reason is that in equilibrium, minute changes by each duopolistic firm are balanced out such that:

(3.1.8) $\boldsymbol{\pi_{i,j}} = \left(p_{1,t}, p_{2,t}, \dots, p_{n,t}\right) \begin{cases} \pi_{i,j}(p_1) & if\ P^{*}{}_1 < P_t \\ \pi_{i,j}(p_2) & if\ P_1 = P_t \\ & 0 \end{cases}$

(3.1.8) is taken from Mayaki (2024) and introduces the payoff discount factor for cooperative actions from Fudenberg and Tirole (1991), which is equal to the following incentive compatibility condition:

(3.1.9) $\lambda K \cdot \frac{1}{1-\delta} \geq \hat{\pi} + \frac{\delta}{1-\delta} \cdot \breve{\pi}$

Table 1.1: *Payoff Matrix*

| Time | Firm *i* | Firm *j* |
|---|---|---|
| *t* | $(\boldsymbol{c_t^i}, \boldsymbol{p_t^i})$ | $(\boldsymbol{c_t^j}, \boldsymbol{p_t^j})$ |
| *t + 1* | $(\boldsymbol{c_{t+1}^i}, \boldsymbol{p_{t+1}^i})$ | $(\boldsymbol{c_{t+1}^j}, \boldsymbol{p_{t+1}^j})$ |
| *T* | $(\boldsymbol{c_T^i}, \boldsymbol{p_T^i})$ | $(\boldsymbol{c_T^j}, \boldsymbol{p_T^j})$ |

Güth *et al.* (2016:69)

## 4. Simulating Growth with Incomplete Information

### *4.1 Defining the Mathematical Model*

This reseach study begins with a nuanced introduction outlining how competition authorities and regulators grapple with the challenge of tacitly collusive pricing behaviour. Such practices—particularly in the era of advanced technology such as AGI—can go undetected. This section focuses on Pareto-optimal behavior in a conglomerate market. In this model of Güth's (2016) social dilemma, there is also an optional export market accessed under an arbitary probability and no profit sharing (Park and Mo, 2014). Revisiting Güth's social dilemma, its possible to argue it thus requires an indefinite level of restraint from firm *i* and firm *j*, respectively, or what Güth (2016) refers to as forbearance, that is, intentional self-restraint from competitive aggression–that is, some sort of tacit cooperation. Herein lies the contemporary challenge of regulation.

These stipulations, while they are not for the faintness of the heart, are, however, precise (3.1.9). Even with an equal proportion of factor endowments and allocations that are at their core, firm *i* may still have more advantageous information and, as such, may grow faster due to qualitative advantages in capital assimilation or what one may call the recursive core (Becker and Chakrabarti, 1995). To observe gains achieved from stable tacit collusion in comparison to gains achieved from defecting or other static outcomes, Suetens and Potters (2007) illustrate what they call a Friedman index (Friedman, 1971). The conditional distribution of a collusive state is said to be optimised when the index is high.

*4.2 Dynamic Stability Analysis*

This section builds on the previous subsection and extends it to explore an area beyond stable optimisation progressing into the important aspect of core allocations.

Let $\dot{x}$ and $\dot{y}$ be a system of equations derived from the LHS of (3.0.1) and from equation (3.0.3) respectively. We can form the following pair of nodes:

(4.2.1) $$\begin{cases} \dot{x} = -\alpha x + by \\ \dot{y} = A \cdot \sigma^{(1-\varphi)} - y \end{cases}$$

- where $x$ signifies initial factor endowments.
- and where $y$ denotes the asymmetric growth from the firm's export intensity

The system can be expressed as:

(4.2.2) $$\begin{bmatrix} \dot{x} \\ \dot{y} \end{bmatrix} = \begin{bmatrix} -a & b \\ 0 & -1 \end{bmatrix} \cdot \begin{bmatrix} x \\ y \end{bmatrix} + \begin{bmatrix} 0 \\ A \cdot \sigma^{(1-\varphi)} \end{bmatrix}$$

If we assume:

- $a = 0.4$
- $b = 0.2$
- $A = 1$, $\sigma = 1.2$ and $\varphi = 0.4$

The Jacobian of the system is defined as:

(4.2.3) $$J = \begin{bmatrix} -0.4 & 0.2 \\ 0 & -1 \end{bmatrix}$$

Calculating Eigenvalues:

$$\begin{bmatrix} -0.4 - \lambda & 0.2 \\ 0 & -1 - \lambda \end{bmatrix} = (-0.4 - \lambda) \cdot (-1 - \lambda) = 0$$

(4.2.4) $$\lambda_1 = -0.4 \text{ and} \lambda_2 = -1.0$$

Both eigenvalues are negative, indicating a stable node (trajectories converge to a stable equilibrium). We now map an N × M phase space around a saddle path to portray the direction of the equilibrium over time, as shown in Figure 1.1.

Figure 1.1: *Phase diagram for N × M interaction*

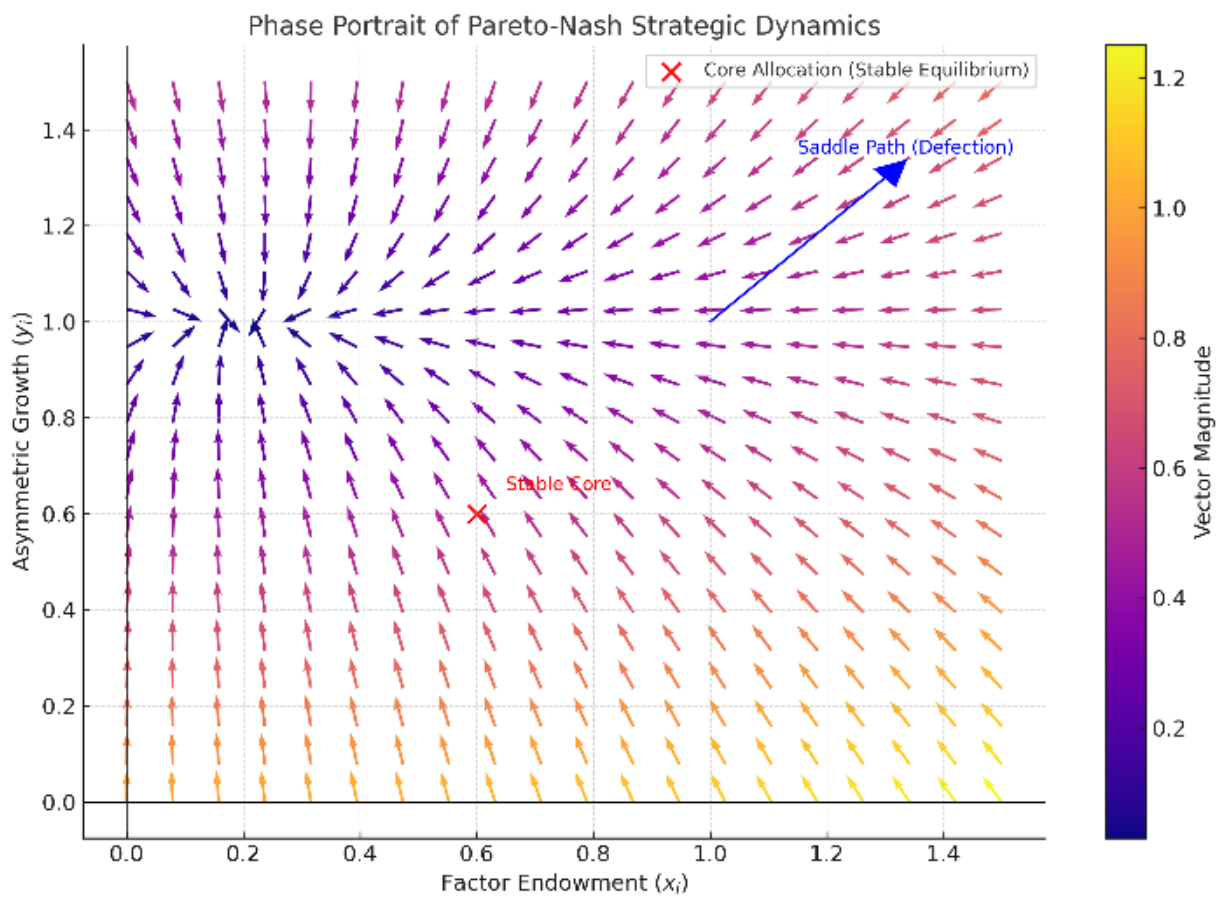

Note the following production function:

(4.2.5) $Y_t = A \cdot W^{\alpha}{}_{i,j,t}, Z^{\omega}{}_{i,j,t}$, where $\alpha + \omega = 1$

As Figure 1.1's phase diagram demonstrates, there is a stable core and saddle path direction. If we now assume that $W_{i,j}$, $Z_{i,j}$ are boundary conditions that are effectual if for all monotonic strategic profiles $W_{i,j}$, $Z_{i,j}$ wherever $\alpha, \omega \neq 0.5$, there is diminishing utility observed by firms $i, j$, which may lead to unstable equilibria, as shown through areas outside the indifference curves in Figure 1.2.

These defective allocations are proven by economists as suboptimal or *recursive* opportunity sets. $W_{i,j}$ *and* $Z_{i,j}$ are thus only in their core when Pareto-optimal growth variables exist within them in Cobb–Douglas form.

Figure 1.2: *N × M Indiffernce and contract curves*

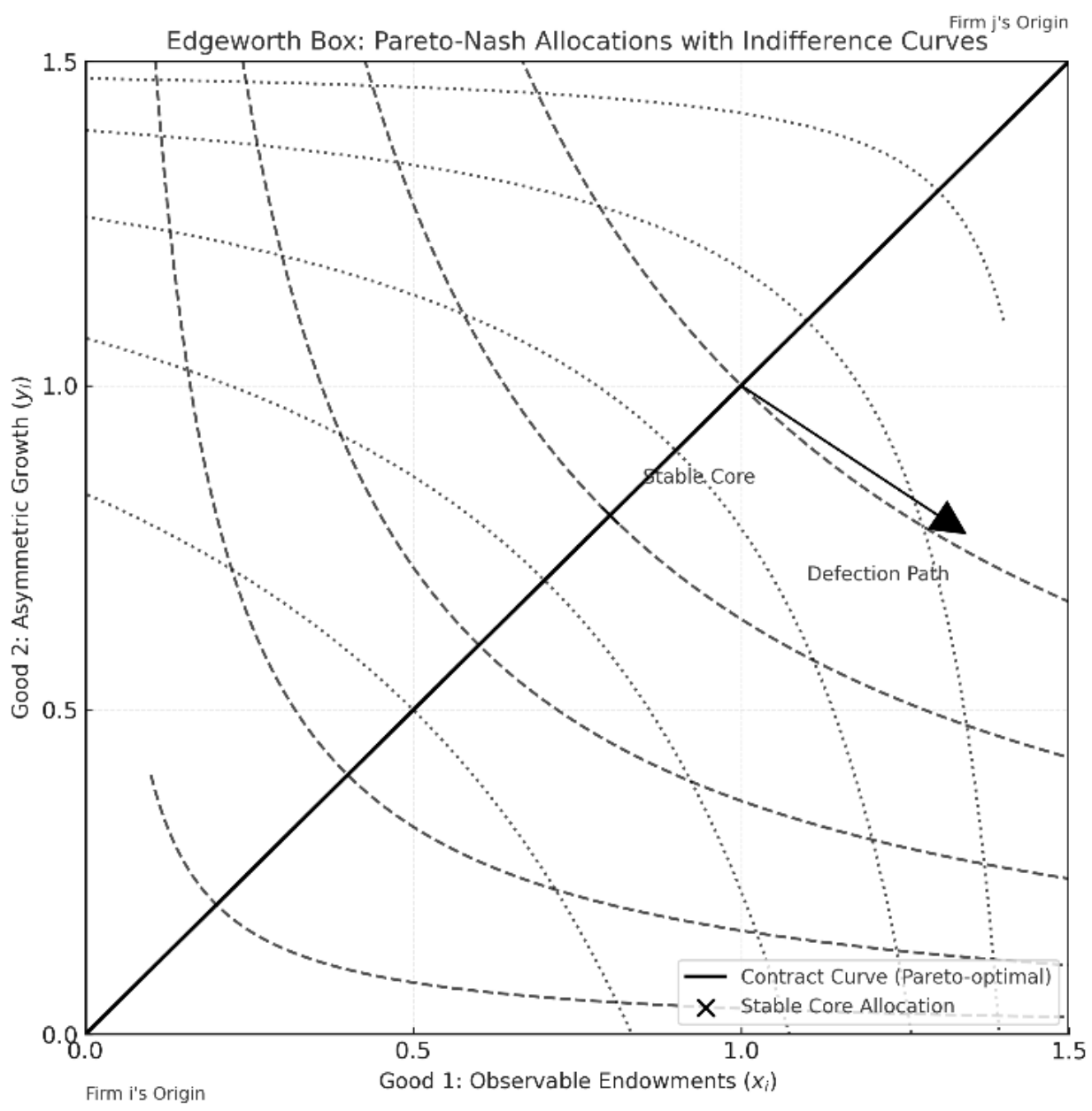

*4.3 Conglomerate Behavior in Price Wars*

Consider a two-stage market involving a differentiated good, denoted as a vector. The market is characterised by a downwards-inclined demand curve, specified by the following function:

(4.3.1) $Y_t = K \cdot {p_t(q_t)}/{1+r}$

Here, $Y_t$ denotes aggregate demand, $K$ denotes the relevant market, and $P_t(q_t)$ represents the composite revenue function for a unit of goods supplied at time $t$ sold to market $K$. $r$ is the discount rate reflecting the intertemporal trade-off in production and consumption.

If we consider an overseas export market of scale $K$, with an idenitical demand curve as found in (4.3.1), and an incumbent profit function defined by (3.0.1), where firm *i* chooses at each stage whether to export to an overseas market, given comparative trade, then firm *i* optimises (3.0.3):

(3.3.2) $\sum_{i=1}^{n}\left(x_{i,t} - A_{i,t} \cdot \sigma_{i,t}^{(1-\varphi)}\right) + \left(\mu_{i,t}p_{i,t} - \beta_{i,t}c_{i,t}\right) \cdot p_{i,t}\left(q_{i,t}\right)$

## 5. Empirical Findings

Now, we investigate the small business case of the duopoly under firm *i*'s first move using a GLM Gaussian identity estimation, where—as per (3.0.1)—the dependent variable for firm $\boldsymbol{i}$ is growth $\boldsymbol{y_{i,t}}$, also known as the regressor—using changes in revenue and headcount over time—shall be regressed with:

(5.0.1) $\hat{y}_{i,t} = \beta_0 + \beta_1 x_{i,t} + \beta_2 A_t + \beta_3 \sigma_{i,t} + \beta_2 \varphi_{i,t} + \varepsilon_{i,t}$

where:

- Initial endowments $x_{i,t}$—reflected by real world balance sheet data
- Firm age $A$—reflected by years of incorporation
- Export intensity $\sigma_{i,t}$—reflected by firm-level trade data
- Information asymmetry $\varphi_{i,t}$—reflected by business survey data
- Residual error term $\varepsilon_{i,t}$ under the null hypothesis of no significance[2].

Table 1.2: Regression Results (GLM – Gaussian Identity)

| | Coefficient | p value | Interpretation |
|---|---|---|---|
| Endowment | 0.5220 | 0.000 | *Strong*, significant *positive effect* on firm growth |
| Firm Age | 0.0050 | 0.698 | Statistically insignificant. Older firms *don't* grow faster, on average |
| Export Intensity | 0.1935 | 0.817 | Not statistically significant |
| Information Asymmetry | 1.4173 | 0.040 | Significant—*less* asymmetry (higher $\boldsymbol{\varphi}$ index) is associated with higher growth |

This finding supports the theoretical intuition that firms with better information access (*lower* asymmetry) grow *more*, even when controlling for size, age and export exposure.

[2] See Addendum 1

The estimated GLM equation for $\boldsymbol{y_{i,t}}$ is as follows:

(5.0.2) $-0.889 + 0.522 \cdot x_{i,t} + 0.005 \cdot A_{i.t} + 0.194 \cdot \sigma_{i,t} + 1.417 \cdot \varphi_{i,t}$

Figure 1.3: *Matplotlib Chart – GLM Regression Results*

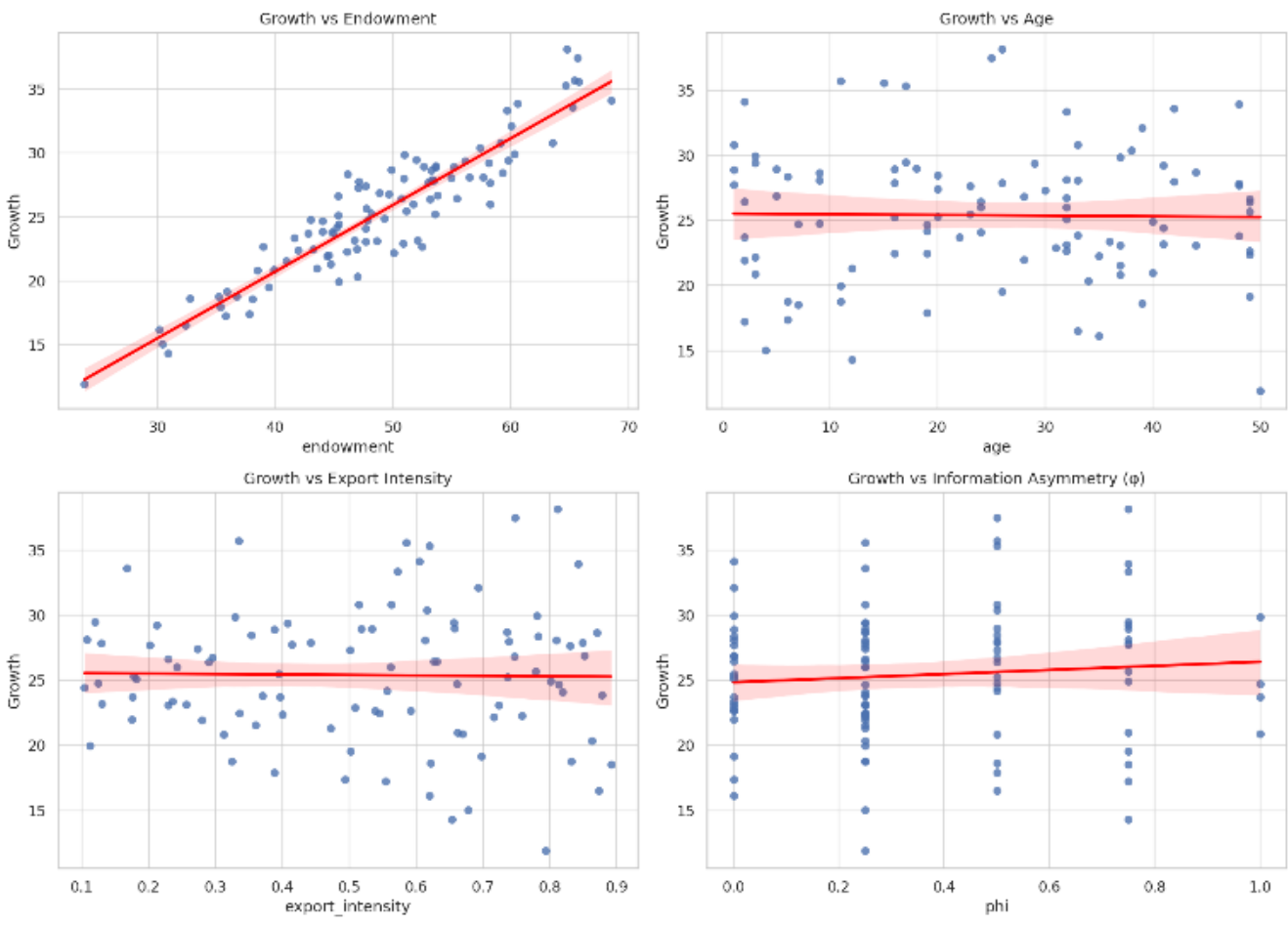

The study will now show results from the same GLM regression model structure taken from the simulation for firm $i$ but recalibrated for an example of a conglomerate. In this case, we reinterpret each variable for a multimarket, multiproduct firm setting and then simulate it via the same index-based GLM approach.

The estimated GLM equation for $\boldsymbol{y_{i,t}}$ now becomes:

(5.0.3) $-0.11 + 0.04 \cdot x_{i,t} + 0.89 \cdot \sigma_{i.t} + 1.00 \cdot \sigma_{i,t} + 0.02 \cdot A_{i,t}$

Finally, this paper reports findings of a study which employs a Fixed Effects (FE) panel regression conducted on a simulated dataset of fifty conglomerate firms observed over five consecutive time periods, intended to represent five fiscal years (2020–2024). This panel structure enables the model to distinguish between within-firm temporal changes and between-firm structural differences, offering a richer basis for analysing firm growth under conditions of incomplete information. By introducing fixed effects for each firm, the model absorbs all unobservable, time-invariant factors such as managerial competence, legacy market positions, and structural integration, which are especially relevant in the case of diversified conglomerates. These firm-specific intercepts ($\alpha_i$) account for baseline growth potential unrelated to year-over-year changes in measured variables.

Within this five-year panel, firm endowment—proxied by log-normalised asset size—was consistently a significant driver of growth. With a coefficient of approximately 0.029 and high statistical significance ($p < 0.001$), the results suggest that even within firms, increases in capital base or operating capacity contribute meaningfully to growth outcomes on an annual basis. This finding is theoretically consistent with the Pareto–Nash framework: well-endowed firms are better positioned to stabilise cooperative outcomes over repeated interactions, especially when strategic forbearance yields advantages that are only visible across multiple periods.

Export intensity, measured as the proportion of firm sales derived from international markets, fluctuated modestly across time for each firm. However, in the fixed effects specification, it returned a negative coefficient (-2.14) and was not statistically significant ($p \approx 0.245$). This contrast with earlier GLM results suggests that the growth advantage of exporters is likely driven by between-firm differences rather than within-firm export variation year over year. In other words, the fixed effects model—by eliminating cross-firm variation—may mask the broader role of global market exposure. Moreover, it is plausible that any short-term gains from expanding the export share may be counterbalanced by increased operational complexity, regulatory exposure, or strategic noise across markets, especially when modelled on an annual horizon.

The information asymmetry index (φ) was constructed as a reverse-scored measure (i.e., higher φ implies greater asymmetry) and varied slightly within firms across the five-year window. Although it plays a foundational role in the theoretical model of cooperative stability, φ was not statistically significant in this FE setting (coefficient ≈ 0.50, $p \approx 0.896$). This is likely due to its relatively low within-firm variance over the short timescale simulated. Over five years, especially in conglomerates, asymmetry may persist structurally owing to governance models, platform complexity, or intermarket opacity—factors that evolve over decades rather than annual increments. Consequently, while φ may be a strong *between-firm* predictor in pooled or random effects models, its explanatory power diminishes when assessing short-term, within-firm dynamics[3].

Firm age, measured in years since incorporation and updated annually, also lacked statistical significance (coefficient ≈ -0.015, $p \approx 0.496$). This outcome reinforces the idea that, over a five-year window, marginal increases in firm age may have little bearing on performance—especially for conglomerates already operating at scale. The absence of a significant age effect also suggests that learning or institutional memory may be fully embedded within the fixed effects term for each firm.

[3] See Addendum 1

Taken together, these results suggest that while the five-year panel captures meaningful within-firm growth dynamics, particularly through changes in endowments, it may not be long enough to capture the strategic evolution of information asymmetry or internationalisation benefits. This reinforces the theoretical insight that stability under incomplete information unfolds over longer horizons and highlights the need for models that combine firm-level persistence (captured via fixed effects) with longer-term dynamics and belief updating mechanisms.

Finally, the estimated fixed effects[4] equation for $\boldsymbol{y_{i,t}}$ is as follows:

(5.0.4) $\alpha_i + 0.029 + x_{i,t} - 2.136 \cdot \sigma_{i.t} + 0.499 \cdot \varphi_{i,t} - 0.015 \ A_{i,t}$

where:

- $\alpha_{i,t}$ is equal to the firm-specific intercept
- $x_{i,t}$ is the endowment size (firm size)
- $\sigma_{i.t}$ is the export ratio
- $\varphi_{i,t}$ is information asymmetry, and
- $A_{i,t}$ is firm age

This is in addition to the residual error term $\varepsilon_{i,t}$.

## 6. Research Implications

### *6.1 Interpreting Regression Results*

The results show that firm endowment matters significantly. In both the GLM and panel regressions, endowment (firm size) was a strong, significant predictor of growth. This finding supports the assumption that resource-abundant firms are better able to sustain forbearance—even when short-term profit maximisation might tempt defection. These firms have the "*slack*" to resist price wars and preserve long-term intermarket stability, aligning with the concept of Pareto efficiency in strategic restraint.

The $\boldsymbol{\varphi}$ index, which captures information asymmetry, played a complex role in the study. In the cross-sectional GLM, $\varphi$ was significantly negative, supporting the null hypothesis that greater asymmetry reduces the capacity for coordinated restraint and thereby lowers long-term growth.

However, in the panel FE model, $\varphi$ lost significance—but this is normal and expected: FE controls for all firm-level constants, so a relatively stable $\boldsymbol{\varphi}$ has little within-firm variation over the five-year sample period.

Thus, the result validates the modelling assumption: incomplete information is a structural determinant of forbearance and Pareto-Nash stability. Its influence is clearest in between-firm comparisons (conglomerates with high internal transparency or digital maturity grow better) but becomes muted when only within-firm year-to-year variation is analysed.

The export ratio was positive and significant in the GLM but negative and insignificant in the fixed effects model. This implies that exporting itself may not foster forbearance—it may even expose firms to more volatility and retaliation risks, depending on the market context. This underscores the fragility of Pareto-Nash outcomes in globally entangled markets, especially under asymmetric visibility.

Furthermore, the fixed effects model revealed that firm-specific characteristics dominate short-run growth outcomes. This finding supports the claim that forbearance is *firm-specific*—not all conglomerates are equally capable of sustaining tacit collusion across markets. The FE framework confirms the relevance of persistent strategic types, which the theoretical model incorporates via the stability condition for Pareto-optimal trajectories in (2.0.9)

The empirical results—especially the GLM and fixed effects estimates—affirm the following:

1. Strategic forbearance is most likely when conglomerates are large, mature, and well informed.

2. Information asymmetry undermines the ability to coordinate and sustain cooperative outcomes, as the mathematical hypothesis model predicts.

3. While FE models reveal the importance of persistent firm-level traits, the GLM confirms that observable asymmetries in information and structure correlate with lower growth—suggesting that policy interventions improving information quality or reducing uncertainty could enable more stable, efficient outcomes in oligopolistic markets.

## 7. Conclusion

This study extends the theoretical framework of Pareto–Nash allocation under incomplete information by focusing on Guth's concept of a social dilemma within a two-stage, multimarket pricing game. By using intuition from accretion disk astrophysics, embedding forbearance as a strategic necessity rather than a behavioural anomaly, we have demonstrated how conglomerate firms, even with asymmetric endowments, may reach stable, core allocations through what we have shown is proof of a cooperative equilibrium. This setup is especially relevant where algorithmic pricing and imperfect information complicate the direct observation of market actions.

[4] See Addendum 2.

Crucially, this model underscores how firms must internalise the trade-off between maximising short-run gains and preserving long-run stability in markets where defection not only guarentees severe retaliative action but also regulatory scrutiny. Firms compete through mutual forbarance—and are hence optimised where all allocations are *in core*—by incorporating conditions for economic certainty and symmetry of information under the jeopardy of profit sharing. In particular, it explores how conglomerate firm behavior and submarket regulation, given an initial market size defined by multiple factor endowments, and how a predetermined growth rate is influenced by price stability in long-run equilibrium. As we have observed, $K$ is quasiexponential, implying that a higher rate of market growth incurs much greater increments of future profitability. The assumption of export intensity has been demonstrated.

In policy terms, this work adds to the growing literature on conglomerate behaviour and regulation, where a Stackleberg leader's Bertrand–Nash cooperation move for entrants in the relevant market resembles collusion with the incumbent, especially when this incumbent is a conglomerate yet produces socially optimal outcomes. The Pareto-Nash framework developed in this study offers both an analytical tool and a policy lens for ensuring competitive but stable markets, which serve the wider objective of price stability.

*End of Document*

## 9. Addends/Appendices

1. Data Sources by Regression Variable (Conglomerate GLM)

| Firm Growth | Firm Size/Endowment | Firm Age | Export Intensity | Information Asymmetry |
|---|---|---|---|---|
| Orbis – Revenue growth, EBITDA | Orbis – Total assets, turnover | Orbis – Incorporation year | Orbis – Exports/sales | BEIS LSBS – digital tools |
| Compustat – Sales growth, Tobin's Q | Compustat – Total assets | OpenCorporates – Company registry | UN Comtrade – Export volumes | BEIS LSBS – strategy/monitoring |
| FAME – Net income growth | FAME – Net assets, equity base | Companies House (UK) | HMRC Trade in Goods by Firm | Understanding Society – uncertainty |
| Bloomberg – EPS growth, ROA | Bloomberg – Book value, revenue | Dun & Bradstreet – metadata | OECD STAN – Export ratios | ONS Innovation Survey – R&D visibility |
| WRDS/CRSP – Stock returns | WRDS – Market cap | UK BSD – firm age bands | Eurostat – Structural stats | WB Enterprise Survey – market access |
| World Bank ES – Sales change | World Bank – Capital stock | Crunchbase – Founding dates | ITC TradeMap – trade flows | IMF FAS – Disclosure index |

2. Fixed Effects Panel Regression Results (Conglomerate Panel & FE)

| Variable | Coefficient | Std. Error | t-Statistic | p-Value |
|---|---|---|---|---|
| endowment | 0.0290 | 0.0008 | 37.52 | 0.000 |
| export_ratio | -2.1364 | 1.8318 | -1.17 | 0.245 |
| phi | 0.4995 | 3.8022 | 0.13 | 0.896 |
| age | -0.0152 | 0.0223 | -0.68 | 0.496 |

3. Mapping Model Concepts to Regression Variables (GLM, Panel and FE)

| Model Concept | Regression Variable | Interpretation |
|---|---|---|
| Firm capacity to absorb losses or maintain strategy | endowment | Firms with more resources can afford to maintain restraint |
| Exposure to retaliatory dynamics | export_ratio | Participation in global markets may encourage or destabilise cooperation |
| Information asymmetry ($\varphi$) | phi_index | Captures uncertainty or opacity that prevents perfect coordination |
| Long-run strategy/firm maturity | age | Tests how experience shapes consistent strategies over time |
| Dynamic firm performance | growth | Used as the outcome measure to evaluate strategic payoff |